\documentclass[aps,prl,twocolumn,superscriptaddress,showpacs]{revtex4}

\usepackage{mathrsfs}
\usepackage{graphicx}
\usepackage{color}
\usepackage{amsmath}
\usepackage{amssymb}

\begin{document}

\title{First passage statistics for aging diffusion in annealed and quenched
disorder}

\author{Henning Kr{\"u}semann}
\affiliation{Institute of Physics \& Astronomy, University of Potsdam, 14776
Potsdam-Golm, Germany}
\author{Alja\v{z} Godec}
\affiliation{Institute of Physics \& Astronomy, University of Potsdam, 14776
Potsdam-Golm, Germany}
\affiliation{National Institute of Chemistry, 1000 Ljubljana, Slovenia}
\author{Ralf Metzler}
\email{rmetzler@uni-potsdam.de}
\affiliation{Institute of Physics \& Astronomy, University of Potsdam, 14776
Potsdam-Golm, Germany}
\affiliation{Department of Physics, Tampere University of Technology, FI-33101
Tampere, Finland}

\date{\today}

\begin{abstract}
Aging, the dependence of the dynamics of a physical process on the time $t_a$
since its original preparation, is observed in systems ranging from
the motion of charge carriers in amorphous semiconductors over the blinking
dynamics of quantum dots to the tracer dispersion in living biological cells.
Here we study the effects of aging on one of the most fundamental properties of a
stochastic process, the first passage dynamics. We find that for an aging
continuous time random walk process the scaling exponent of the density of first
passage times changes twice as the aging progresses and reveals an intermediate
scaling regime. The first passage dynamics depends on $t_a$ differently for
intermediate and strong aging. Similar crossovers are obtained for the first
passage dynamics for a confined and
driven particle. Comparison to the motion of an aged particle in
the quenched trap model with a bias shows excellent agreement with our analytical
findings. Our results demonstrate how first passage measurements can be used to
unravel the age $t_a$ of a physical system.
\end{abstract}

\pacs{72.20.Jv,72.70.+m,89.75.Da,05.40.-a}

\maketitle 

In their groundbreaking 1975 paper Scher and Montroll introduced the scale-free
distribution $\psi(\tau)\simeq\tau^{-1-\alpha}$ with $0<\alpha<1$ of trapping
(waiting)
times $\tau$ for moving charge carriers in amorphous semiconductors to explain
the measured power-law form of the electrical current \cite{scher,neher1}. This then
radical assumption has since then been verified in numerous systems \cite{revs}
including the tracer dispersion in groundwater aquifers \cite{brian}, the motion
of endogenous protein channels and granules in biological cells \cite{tabei,jeon}
and of submicron tracers in structured environments \cite{wong}, or the blinking
dynamics of quantum dots \cite{dots}.

The lack of a time scale $\langle\tau\rangle$ in these systems effects the
disparity between ensemble and time averages of physical observables (weak
ergodicity breaking) \cite{web,pt} and an explicit dependence of observables on
the time span between the initial preparation of the system and start
of the measurement at $t_a$, the so-called aging \cite{age,eli_age,paolo}.
Aging phenomena
can be rationalized by particle dynamics in quenched energy landscapes \cite{qel}
or logistic maps \cite{maps}, and were experimentally observed in biological
systems \cite{tabei} as well as amorphous semiconductors \cite{neher}.

Here we present analytical and numerical evidence for the distinct effects of
aging on the first passage properties of aging systems, i.e., the statistics of
the times when the process first crosses a pre-set value. In the above examples this
is the arrival of charge carriers at the counterelectrode, giving rise to the
decay of the electrical current, the breakthrough of chemical tracers at some
probe location in an aquifer, the arrival of a protein channel at a specific
receptor in a cell membrane, or for a quantum dot to reach its $n$th on-state. Our
results for the first passage time density (FPTD) $\wp(t)$ in systems with a
diverging
time scale $\langle\tau\rangle$ exhibit a clear dependence on the aging time
$t_a$ in the cases of intermediate and strong aging while $\wp(t)$ is
independent of $t_a$ for short aging times. More importantly, progressing
aging changes the scaling exponent of $\wp(t)$ and reveals an interesting
intermediate scaling regime. For biased diffusion we show
that our results agree well with simulations in a quenched energy landscape.
Finally we address the question how first passage measurements can be used to
unravel information on the age $t_a$ of the observed system.

In the absence of aging ($t_a=0$), i.e., when the measurement commences
simultaneously with the initiation of the system at $t=0$, the Scher-Montroll
continuous time random walk (CTRW) process with trapping time density $\psi(\tau)
\simeq\tau^{-1-\alpha}$ and $0<\alpha<1$ causes free anomalous diffusion of the
subdiffusive form
$\langle x^2(t)\rangle\simeq K_{\alpha}t^{\alpha}$, with the anomalous diffusion
coefficient $K_{\alpha}$ \cite{scher,revs}. On the semi-axis with $\delta$-initial
condition at $x=0$ the FPTD to the point $x_0$ associated with this CTRW process
reads
\begin{equation}
\label{nonFPD}
\wp(t)=\left(\frac{K_{\alpha}}{x_0^2}\right)^{1/\alpha}l_{\alpha/2}\left(\left[
\frac{K_{\alpha}}{x_0^2}\right]^{1/\alpha}t\right)\simeq\frac{x_0/K_{\alpha}^{1/2}
}{t^{1+\alpha/2}}
\end{equation}
in terms of the one-sided L{\'e}vy stable law $l_{\alpha/2}(t)$ \cite{hughes}. Its
asymptotic expansion shows the power-law decay $t^{-1-\alpha/2}$ characteristic
for unbiased subdiffusion \cite{revs}. Eq.~(\ref{nonFPD}) follows from the Laplace
image \cite{laplace} $\wp(u)=\exp(-x_0u^{\alpha/2}K_{\alpha}^{1/2})$ derived in
Ref.~\cite{bvp}. In the Brownian limit $\alpha=1$, the FPTD reduces to the familiar
L{\'e}vy-Smirnov law with asymptote $\wp(t)\simeq x_0/[K_1t^{3/2}]$ exhibiting the
famed Sparre-Andersen $3/2$ universality \cite{redner}.

To extend this result for an aged system ($t_a>0$) we use the propagator described
by the aging CTRW \cite{eli_age,johannes}
\begin{equation}
\label{prop}
P_a(k,s,u)=P_0(s,u)+h(s,u)[u+K_{\alpha}u^{1-\alpha}k^2]^{-1}
\end{equation}
in Fourier-double Laplace representation $P_a(x,t_a,t)\to P(k,s,u)$ \cite{fourier}.
The major component in the mathematical description of aging processes is the
density $h(t_a,t)$ for the so-called forward waiting time $t$ for the occurrence
of the first step in the random walk process, after the system aged for $t_a$. Due
to the lack of a characteristic scale of $\psi$ increasingly longer trapping times
occur while the system evolves. Typically, after the aging period the system is
arrested in such a long trapping state, thus changing the statistics of the first
step to occur, as given by the distribution $h$. Its
double Laplace transform is $h(s,u)=[\psi(s)-\psi(u)]/([u-s][1-\psi(s)]$
\cite{eli_age,johannes}. In Eq.~(\ref{prop}) $P_0(s,u)=[1-sh(s,u)]/(su)$ is the
Laplace transform of the probability $P_0(t_a,t)$ that no step occurs up to time
$t$. The splitting into a discrete part for completely immobile particles and a
continuous portion weighted by the density $h$ is typical for aging CTRW processes
\cite{johannes}.

\begin{figure}
\includegraphics[width=8.8cm]{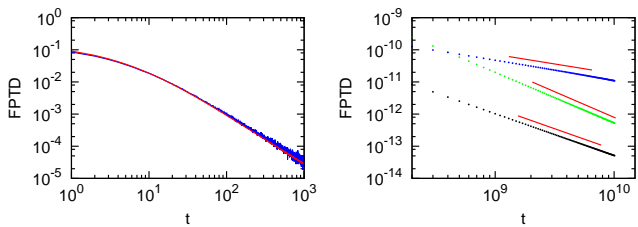}
\caption{Semi-infinite domain. Left: Simulation (blue) and analytical result (red)
for $\alpha=0.75$, $t_a=100$, $x_0=1$, and $K_{\alpha}=0.5$. Right: Different
scaling regimes in the long time limit for $t_a=0.1$ (black, bottom), $t_a=10^7$
(green, middle), and $t_a=10^{11}$ (blue, top), with $\alpha=0.6$, $x_0=1$, $K_{
\alpha}=0.2$.}
\label{semiinf}
\end{figure}

Using the standard method of images \cite{redner} and the subordination trick
\cite{revs,igors} we find the general closed form
\begin{equation}
\label{conv}
\wp_a(t_a,t)=\left(\frac{K_{\alpha}}{x_0^2}\right)^{1/\alpha}h(t_a,t)\otimes
l_{\alpha/2}\left(\left(\frac{K_{\alpha}}{x_0^2}\right)^{1/\alpha}t\right)
\end{equation}
for the FPTD, where $\otimes$ denotes a Laplace convolution. This form is very
useful for numerical evaluation. The exact expression for $\wp_a(t_a,t)$ involves
an infinite power series \cite{long}. In the long
time limit ($t\to\infty$) we find the three scaling regimes
\begin{equation}
\label{semi_scale}
\wp_a(t_a,t)\simeq\left\{\begin{array}{ll}t_a^{\alpha-1}t^{-\alpha},&t_a\gg t\\
t_a^{\alpha}t^{-1-\alpha},&t_a\ll t\ll t^2_a\\
x_0K_{\alpha}^{-1/2}t^{-1-\alpha/2},&t^2_a\ll t
\end{array}
\right.
\end{equation}
depending on the severity of the aging. Thus, when aging is weak, the asymptotic
behavior of the non-aged process from Eq.~(\ref{nonFPD}) is preserved. Remarkably,
once the aging becomes more pronounced, the competition between the magnitudes of
the measurement time $t$ and the aging time $t_a$ effects a change of the scaling
exponent of $t$ from $1+\alpha/2$ to $1+\alpha$ at intermediate values of $t_a$,
and a further change to $\alpha$ under strong aging conditions. The crossover
between these scaling regimes is a key signature of scale-free CTRW processes.
Fig.~\ref{semiinf} on the left demonstrates excellent agreement of our analytical
result (\ref{conv}) with simulations of the CTRW process with trapping time density
$\psi(\tau)$. On the right of Fig.~\ref{semiinf} we confirm the existence of the
three different scaling regimes of $\wp(t)$ predicted by Eq.~(\ref{semi_scale}),
again observing excellent agreement. Note, however, that in order to see all
three regimes, the variation of $t_a$ needs to be quite large. Thus, depending on
the physical system and the experimental technique not all three regimes may be
detectable. As the first crossover in Eq.~(\ref{semi_scale}) increases the
magnitude of the scaling exponent while the second crossover decreases it again,
the FPTD behavior nevertheless provides a new method to deduce the age
$t_a$ of an aging system. The discovery of three distinct scaling regimes and the
dependence of the FPTD on the aging time $t_a$ are our first main results.

\begin{figure}
\includegraphics[width=8.8cm]{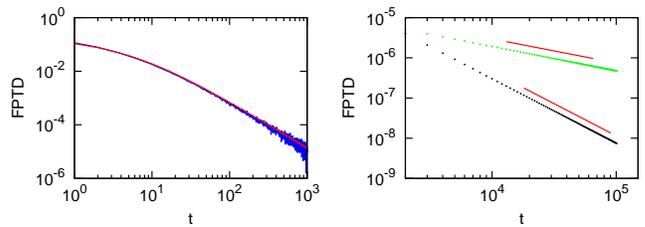}
\caption{Finite domain. Left: Simulation (blue) and analytical result (red)
for $\alpha=0.75$, $t_a=100$, $x_0=1$, and $K_{\alpha}=0.5$. Right: Different
scaling regimes in the long time limit for $t_a=0.1$ (black, bottom) and
$t_a=10^7$ (green, top), with $\alpha=0.6$, $x_0=1$, $K_{\alpha}=0.2$.}
\label{doubleR}
\end{figure}

\emph{Finite domain.} For the first passage to the boundaries of a finite domain
the asymptotic scaling of the non-aged FPTD is $\wp(t)\simeq x_0^2K_{\alpha}^{-1}
t^{-1-\alpha}$ \cite{bvp}, i.e., the decay is steeper than in the semi-infinite
case. Due to the divergence of the trapping time scale $\langle\tau\rangle$,
however, the mean first passage time still diverges, in contrast to the Brownian
case ($\alpha=1$), for which $\wp(t)$ has an exponential cutoff \cite{redner}.
For the aged system, we again derive $\wp_a(t_a,t)$ via the images method. From a
convolution similar to Eq.~(\ref{conv}) we find an exact solution in terms of a
power series with Lerch functions \cite{long}. The long-time scaling
\begin{equation}
\label{finite}
\wp_a(t_a,t)\simeq\left\{\begin{array}{ll}t_a^{\alpha-1}t^{-\alpha},&t_a\gg t\\
\mbox{\boldmath$($}x_0^2/[2K_{\alpha}]+t_a^{\alpha}/\Gamma(1+\alpha)\mbox{\boldmath
$)$}t^{-1-\alpha},&t_a\ll t\end{array}\right.
\end{equation}
emerges, where this time we only observe two scaling regimes in the measurement
time $t$: at weak aging, the scaling exponent is $1+\alpha$, which changes to
$\alpha$ at strong aging. Concurrently the aging time does not appear explicitly
as long as $t_a\ll\mbox{\boldmath$($}x_0^2\Gamma(1+\alpha)/[2K_{\alpha}]\mbox{
\boldmath$)$}^{1/\alpha}$. At intermediate aging, the prefactor $t_a^{-\alpha}$
enters, while for strong aging it changes to $t_a^{-1-\alpha}$. Fig.~\ref{doubleR}
shows excellent agreement with these results. The behavior of the aged FPTD on a
finite domain is our second important result.

\begin{figure}
\includegraphics[width=8.8cm]{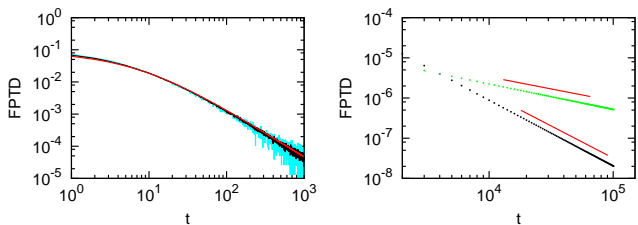}
\caption{Semi-infinite domain, biased case. Left: Simulations of quenched trap
model (cyan) and CTRW (black), and analytical result (red), for $\alpha=0.5$, $t_a=
100$, $F=1$, $T=0.5$, $x_0=0.2$, $\tau_0=10^{-4}$. Right: Different
scaling regimes for $t_1=0.1$ (black, bottom) and $t_a=10^7$ (green, top), for
$\alpha=0.6$, $x_0=1$ and $K_{\alpha}=0.2$.}
\label{regimeF}
\end{figure}

\emph{Biased diffusion.} In many physical systems the motion of the particle is
biased by an external force, for instance, the electrical field acting on the
charge carriers in the amorphous semiconductor of Ref.~\cite{scher} or a subsurface
water stream dragging along the dissolved tracer chemicals in groundwater aquifers
\cite{brian}. To explore the effects of an external bias on the FPTD we now add the
force $F$ to the dynamics. In the classical Brownian case, a bias towards the
absorbing boundary leads to an exponential decay of the associated FPTD, with the
mean first passage time $x_0m\eta/F$, where $\eta$ is the friction coefficient.
In the non-aged case with a scale-free distribution $\psi(\tau)$ of trapping 
times, the FPTD has the power-law form $\wp(t)\simeq x_0t^{-1-\alpha}$ and the
mean first passage time diverges. When the system is aged, the method of images
with an appropriate correction factor still applies \cite{redner} and we
find the FPTD
\begin{equation}
\label{biased}
\wp_a(t_a,t)=h(t_a,t)\otimes\left\{\otimes_{k=1}^{\infty} g_k(t)\right\}
\end{equation}
as a multiple convolution $\otimes_{k=1}^ng_k(t)=g_1\otimes g_2\otimes\cdots
\otimes g_n(t)$ of the function $g_k(t)=\mathcal{D}(\alpha,k)l_{k\alpha}
\mbox{\boldmath$($}\mathcal{D}(\alpha,k)t\mbox{\boldmath$)$}$ with \cite{REM}
\begin{equation}
\label{aux}
\mathcal{D}(\alpha,k)=\left[ \left(\frac{2T}{F}\right)^{2k-1}\left(\frac{2}{K_{
\alpha}}\right)^{k}\frac{\Gamma(3/2)}{\Gamma(3/2-k)k!}\right]^{-1/k\alpha}.
\end{equation}
Here, the temperature $T$ enters due to the competition between the force $F$
and the thermal energy $k_BT$ included through the generalized Einstein-Stokes
relation $K_{\alpha}=k_BT/(m\eta_{\alpha})$, where $\eta_{\alpha}$ is the
generalized friction coefficient \cite{prl}. The exact solution involves a double
series with generalized regularized hypergeometric functions \cite{long}, from which
two scaling forms can be distinguished in the limit of long measurement times,
\begin{equation}
\wp_a(t_a,t)\sim\left\{\begin{array}{ll}t_a^{\alpha-1}t^{-\alpha},&t_a\gg t\\
\left(x_0T/[FK_{\alpha}]+t_a^{\alpha}/\Gamma(1+\alpha)\right)t^{-1-\alpha},&t_a
\ll t\end{array}\right..
\end{equation}
Similar to the result (\ref{finite}) for the first passage in a finite domain we
obtain the crossover from the measurement time scaling with exponent $1+\alpha$ to
$\alpha$ with increased aging, while the aging time appears explicitly for
intermediate and strong aging. Fig.~\ref{regimeF} shows excellent agreement of
our exact result with simulations.

\begin{figure}
\includegraphics[width=8.8cm]{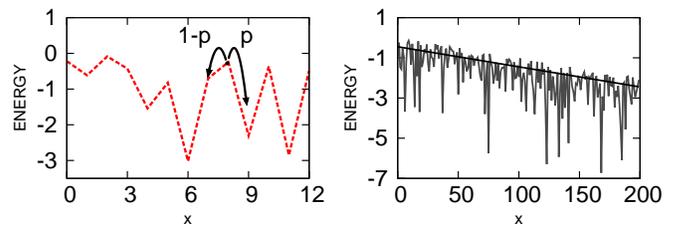}
\caption{Left: Schematic motion in quenched trap model with energy-dependent
transition probabilities $1-p$ and $p$ for moving to the left and right. Right:
Realization of the tilted quenched trapped model with
$T_g=1$, $F=1$, and $a=10^{-3}$. The lattice constant in this Figure is 2,
while we chose $0.01$ in the simulations for Fig.~\ref{regimeF}.}
\label{landscape}
\end{figure}

\emph{Quenched trap model.} The quenched trap model \cite{qel} is often used as
a physical model for CTRW processes with scale-free trapping times. In this lattice
model each site is assigned a random energy value, taken from an exponential
density $p(E)=T_g^{-1}\exp(-E/T_g)$, where we set the Boltzmann constant to unity.
$T_g$ is the system-specific `glass' temperature setting the scale for the energy
distribution $p(E)$ \cite{qel}.
A sample realization for such a quenched energy landscape is displayed in
Fig.~\ref{landscape}. At lattice site $x$ the walker faces the trap energy
$E_x$ and needs to escape by thermal fluctuations (Kramers escape). After escaping
a trap, the walker jumps to one of the two nearest sites and is trapped again, see
the schematic in Fig.~\ref{landscape}. According to the Arrhenius law the trapping
time at $x$ becomes $\tau_x=\tau_0\exp(E_x/T)$, where $T$ is the bath temperature
and $\tau_0$ is an inverse microscopic rate of escape
attempts. The combination of the density $p(E)$ with the Arrhenius law yields the
long-tailed distribution of trapping times, $\psi(\tau)=\mu\tau_0^{\mu}\tau^{-1-
\mu}$ with the scaling exponent $\mu=T/T_g$. When $T<T_g$, the quenched trap
model thus leads to a power-law trapping time density with diverging characteristic
trapping time $\langle\tau\rangle$ \cite{qel}.

In the CTRW model individual trapping times $\tau$ all have the same distribution
$\psi$ but are independent variables, such that the system is renewed each time a
new trapping time is drawn from $\psi$. This annealed form of the disorder contrasts
the quenched nature of the trap model. A walker that leaves a trap may revisit it
during later jumps, and the ensuing random walk thus exhibits correlations and is
no renewal process. Such correlations are irrelevant in dimensions three and higher,
as random walks are transient. To avoid such correlations in lower dimensions we
equip the quenched landscape with the external bias force $F$ \cite{burov2012weak}.
This force tilts the quenched trap landscape (Fig.~\ref{landscape}) and thus
minimizes the likelihood that the walker returns to previously visited sites
\cite{qel}.

Our simulations included an initial aging routine, in which the walker moves
along the lattice in a given realization of the quenched energy landscape with
the additional bias due to the external force $F$. After the aging for the
period $t_a$, an absorbing site is introduced, which is placed at a the distance
$x_0$ from the walker, independent of its initial position on the lattice. The
resulting first passage time is recorded, and this computer experiment averaged
over $10^5$ realizations of the quenched landscape.

The result for the FPTD obtained from the simulations of the quenched trap model
matches excellently with both the exact result and the simulation of the annealed
CTRW model, as shown in Fig.~\ref{regimeF}. We thus expect that the first passage
dynamics of an aged particle to a surface in unbiased quenched trap landscapes in
three dimensions can be described by results (\ref{semi_scale}) or
(\ref{finite}) depending on whether the volume is finite or semi-infinite.

\emph{Conclusion.}
In this paper we studied the first passage dynamics of an aging diffusion process.
In the CTRW model with scale-free trapping time distribution we showed how aging
changes the scaling exponent of the measurement time $t$ and how the aging time
$t_a$ appears in the FPTD. In the semi-infinite case we revealed three distinct
scaling regimes, while on a finite domain and in the biased case two scaling
regimes appear. Our exact results were demonstrated to agree perfectly with
simulations of the CTRW process. In the biased case, the first passage dynamics
measured in the quenched trap model showed excellent agreement with the CTRW
approach.

A general feature observed in our results is that aging significantly reduces
the efficiency of the first passage, as seen by the decrease of the slope of
the FPTD in an aged system. This reflects a general property of processes with
scale-free trapping times: as the system evolves, longer and longer trapping
times appear on average and lead to a slowing-down of the dynamics. For free
subdiffusion this corresponds to an effectively time-dependent diffusivity
$K_{\mathrm{eff}}\simeq t^{\alpha-1}$. In the quenched trap model this would
correspond to
the particle finding an ever deeper trap during its motion across the energy
landscape.

The first passage time represents one of the most fundamental concepts in
stochastic processes. In many systems it is fairly easy to experimentally or
numerically record the first passage dynamics. Our findings presented here for
aged systems complement the classical results for the first passage dynamics in
non-aged systems governed by scale-free trapping times. In condensed matter systems
such as amorphous semiconductors aging is a relevant concern for applications, as
it significantly changes the first passage dynamics and thus the signatures
of the electrical current. As the age of amorphous semiconductors can be reliably
controlled in experiment \cite{neher}, such systems would be ideal to further test
the CTRW model for charge carrier transport. In groundwater systems governed by
scale-free trapping time distributions,
the reduced first passage efficiency would imply an increased retention of
potentially detrimental chemicals dissolved in the water. And, finally, in
biological systems, aging effects and the associated population splitting lead
to the progressive immobilization of particles with potentially relevant
biological function \cite{tabei,pt,johannes,diego}.

The lack of a characteristic trapping time of both the annealed aging CTRW process
or the motion in the quenched energy landscape naturally makes the process
non-stationary, a property, that in turn is closely related to the non-ergodicity
of the system \cite{pnas}. Using relatively simple experimental methods to probe
the first passage statistics of a system, such as the charge carrier transport in
amorphous semiconductors, as function of measurement and aging times would
represent a direct way to determine the non-stationarity of a system.

With the rapid advance of single molecule tracking techniques it has become
possible to diagnose experimentally recorded time series from individual
particle trajectories with respect to the very stochastic mechanism behind the
measured anomalous diffusion by various complementary tools \cite{pccp,pvar,igor,
fpt,radons,saxton}.
This information of the nature of some particle's dynamics, in particular,
whether it is an ergodic or non-ergodic motion, then allows one to deduce important
consequences for the systems such as the (ir)reproducibility of experiments or the
dynamics of followup processes like the diffusion-limitation of reactions. With
the characteristic crossovers between different scaling regimes of the measurement
time $t$ and the explicit aging time dependence the current results provide a
powerful additional tool to probe the underlying stochastic mechanism. Moreover,
given a sufficiently wide measurement window our results allow to read out the
actual age $t_a$ of the system from measured first passage data. This would be
an extremely useful information for systems with unknown age.

AG acknowledges funding through an Alexander von Humboldt Fellowship. RM
acknowledges funding from the Academy of Finland (FiDiPro scheme).


\begin{thebibliography}{99}
\bibitem{laplace} The Laplace transform of a function $f(t)$ is defined as
\[
f(u)=\mathscr{L}\{f(t);t\to u\}=\int_0^{\infty}f(t)\exp(-ut)dt.
\]
\bibitem{fourier} This transform is defined in terms of
\[
P_a(k,s,u)=\mathscr{L}\left\{\mathscr{L}\left\{\int_{-\infty}^{\infty}P_a(x,t_a,t)
e^{ikx}dx;t_a\to s\right\};t\to u\right\}.
\]
\bibitem{REM} Note that we use the symbol $l_{k\alpha}(z)$ here for a function,
that is formally defined in terms of its Laplace image as $l_{k\alpha}(u)=\exp
(-u^{k\alpha})$. When $k\alpha>1$, this is no longer a one-sided stable law.
\end{thebibliography}
\end{document}